\documentstyle[graphicx]{elsart}

\begin{document}

\begin{frontmatter}

\title{Monster Redshift Surveys through Dispersive Slitless Imaging: 
The Baryon Oscillation Probe}     

\author{Karl Glazebrook, Ivan Baldry, Warren Moos,}
\author{Jeff Kruk, Stephan McCandliss}

\address{Dept.\ of Physics \& Astronomy, Johns Hopkins Univ.,
Baltimore, MD 21218, USA}

\begin{abstract}
  Wide-field imaging from space should not forget the dispersive dimension.
  We consider the capability of space-based imaging with a slitless grism:
  because of the low near-infrared background in space and the high
  sky-density of high redshift emission line galaxies this makes for a very
  powerful redshift machine with no moving parts. A small 1m space telescope
  with a 0.5 degree field of view could measure redshifts for $10^7$ galaxies at $0.5<z<2$
  per year, this is a MIDEX class concept which we have dubbed `The Baryon
  Oscillation Probe' as the primary science case would be constraining dark
  energy evolution via measurement of the baryonic oscillations in the galaxy
  power spectrum. These ideas are generalizable to other missions such as SNAP
  and DESTINY.
\end{abstract}
\end{frontmatter}

\section{Dispersive Space Imaging}

\def\Ha{H$\alpha$} 
\def\micron{$\mu$m} 

Slitless dispersive imaging has a long history extending back to the early
Schmidt plate objective prism surveys. The set up is an imaging system, a
dispersing element (ideally located at a pupil of the imaging system to avoid
aberrations, but this is not essential in slow beams) and a filter to delimit
the bandpass.  In essence the object forms its own slit due to its angular
size, to first order the background over the field of view is uniform and is
equal to the undispersed background through the same filter. From the ground
this means the background remains bright when the object's light is dispersed
and so one takes an enormous signal-to-noise (S/N) penalty.  Because of this
such surveys have generally been restricted to bright emission line galaxies
in the local Universe (see the KISS survey\cite{Weg03} for a recent example).
For unresolved emission lines the S/N is independent of the spectral
resolution of the disperser.

A low background environment greatly increases the sensitivity of slitless
surveys. In space the background is dominated by solar scattered light from
zodiacal dust and at 1.6 microns is $\sim 1000\times$ darker than on the
ground \cite{JWST} (there is some dependence on ecliptic latitude).  On the
Hubble Space Telescope both the ACS optical camera and the NICMOS
near-infrared camera are equipped with grisms for slitless spectroscopy. This
mode has been used on ACS for spectroscopy of Type 1a supernovae to
$z=2$ \cite{Riess04} and on NICMOS for detection of an abundant population of
$z>1$ \Ha\ emission line galaxies \cite{Mc99}. The strong evolution of the
cosmological star-formation rate out to $z=1$ means that their
are many more bright emission line galaxies than would otherwise be the case\cite{Hop00}.
The space background peaks at 0.6\micron\ and so as one follows \Ha\ to
high-redshift both the strong evolution in the source population and the
diminishing background act to counteract the cosmological dimming for the S/N.
Going from $z=0.2$ to $z=2$ these effects work to give us a S/N boost of a
factor of 30 and make the high-redshift regime accessible to small space
telescopes.
 
 \section{The Baryon Oscillation Probe}
 
 During discussions at Johns Hopkins in 2003--2004 we developed a strawman
 concept for a MIDEX class space mission which could do large scale slitless
 redshift surveys. We called this the `Baryon Oscillation Probe' as our
 primary science motivation was to probe the baryonic oscillations in the
 power spectrum of galaxy clustering (see below). Our strawman concept
 consisted of a 1m telescope with a 0.5$^\circ$ FOV operating in the
 1--2\micron\ near-infrared bandpass. This field would be sampled with 0.5
 arcsec pixels giving a $4096 \times 4096$ detector array requirement. A set
 of low-dispersion fixed grisms and filters would provide for dispersed
 slitless imaging. Our rough order-of-magnitude estimates of size, mass and
 cost put this spacecraft in the MIDEX class.  Initial studies show
that BOP could be done either in LEO (with active cooling) or at L2
(passive cooling). We use this strawman to demonstrate
 the power of the slitless concept.

 The exposure times depend on redshift, the response of the system and the
 choice of blocking filters.  Cosmological dimming makes the highest redshifts
 more difficult. Our baseline goal was to measure \Ha\ over the redshift range
 $0.5<z<2$, this is the redshift range optimal for testing dark energy models
 \cite{LH03}. To determine exposure times we use published luminosity
 functions for \Ha\ in this redshift range \cite{Hop00}:
\begin{equation}
  [\log (L^{*}/{\rm W}),\, \log(\phi^{*}/{\rm Mpc^{-3}}),\, \alpha] = [35.9,\,
  -3.1,\, -1.6];
\end{equation}
 and set a luminosity limit as a function of redshift which gives critical
 number density sampling of the galaxy power spectrum for the purposes of the
 baryon oscillation science. We assumed the luminosity function $L^*$ evolved
 as $(1+z)^3$ over $0<z<1$ and was constant for $z>1$. This is a good match to
 the observed data of the cosmological star-formation rate evolution
 \cite{Glz03}.  Following \cite{BG03} we adopt a density of galaxies $n$ sufficient
 that the power spectrum $P(k)$ measurement is not shot-noise limited ($nP=3$
in their criteria) and also a bias model for
 galaxies which evolves with redshift in order to match observations of low
 and high-redshift galaxy clustering ($r_0 = 5 h^{-1}$ Mpc $\simeq$ const.).
 This gives an \Ha\ flux limit which is a function of redshift reaching
 $10^{-16}$ ergs cm$^{-2}$ s$^{-1}$ and a number density of 2000 deg$^{-2}$
 for the $1.75<z<2$ bin. As a cross-check on our luminosity function
 calculation we can count directly the number of sources in redshift bins vs.\ 
 flux in the NICMOS grism fields\cite{Hop00} and find agreement with our numbers.

\begin{figure}[htbp]
\begin{center}
  \includegraphics[width=13cm]{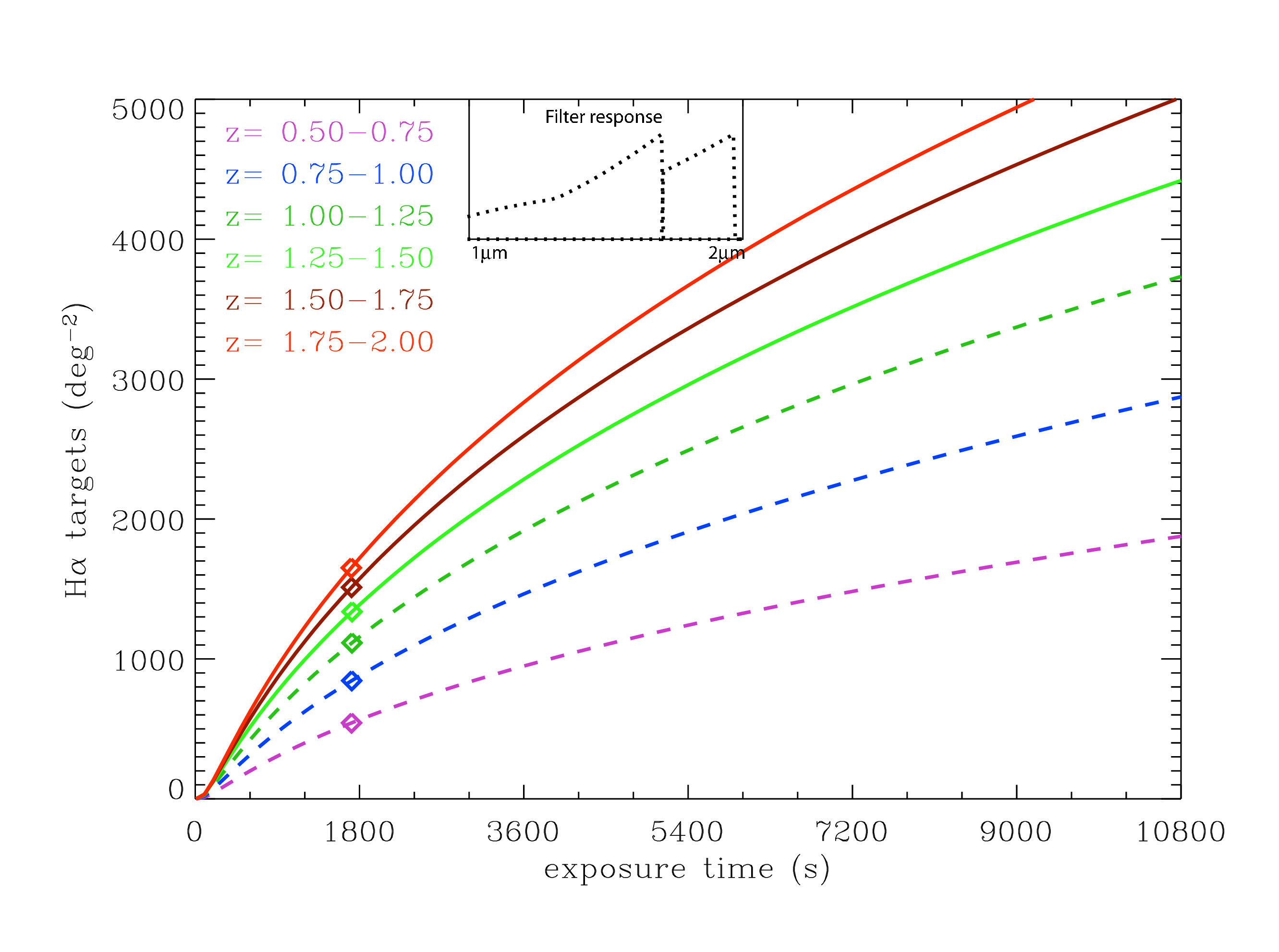}
  \caption{Predicted  number density of targets 
    seen from BOP vs.\ exposure times for different redshift ranges using an
    optimized two filter set. The diamonds show the required number density
    for the baryon oscillation science at each redshift.  The filters have
    been optimized to give equal exposure times for these points. The
    throughputs ($+$ grism response) of the optimized filter set vs.\ 
    wavelength are shown in the inset. We assume a peak efficiency of 38\%
    (effective area of 0.3\,m$^2$) for first-order diffraction and that other
    orders do not contribute significantly to the background.}
\label{fig:filt}
\end{center}
\end{figure}

 Given these flux limits the main design parameter affecting exposure times is
 the combined throughput vs.\ wavelength of the grism and blocking filter. A
 wider filter lets through \Ha\ over a wider redshift range but also lets in
 more background which reduces S/N. Because the background spectrum is not
 flat it is advantageous to have a set of narrow filters rather than one large
 filter. Further the background is {\em higher at shorter wavelengths, i.e.
   low redshifts where the required flux limit is less deep.} This turns out
 to be critical --- if we have the freedom to have an {\em abitrary} response 
function  we find that an {\em optimized\/} filter would
 deliberately have a declining throughput at blue wavelengths. This reduces
 the background, benefiting long-wavelength high-redshift \Ha. The sensitivity
 is reduced at low redshift but the lines are brighter so it does not matter.
 Figure~\ref{fig:filt} shows a hypothetical two filter set designed in this manner
 which would  give a {\em
   uniform} exposure time (1800 secs) for the whole $0.5<z<2$ range. 
   We have not explored the feasibility of such filters, we note that to achieve
   the desired behavior one could change both the filter response or the grism response
   or a combination of both.
   
One issue for slitless surveys is ambiguous line identification. In principle
multi-color photometry can be used to estimate redshifts and remove this
ambiguity. However it is interesting to consider how well one can do with a
single line. Using our evolving line luminosity function model we can estimate
this by adopting reasonable values for fixed emission line ratios and assuming
all the lines following the same cosmological evolution. We find that at our
required flux limits \Ha\ dominates, for example at $z=1.9$ it contributes
72\% of the sources. The principal contaminant turns out to be [SIII] 9532\AA\ 
which aliases \Ha\ at $z=1.1$ (27\% of line sources). Even very crude
broad-band data would suffice to eliminate such low-redshift galaxies. For
[OIII] at $z=2.8$ and [OII] at $z=4.1$ we find contamination rates of 0.5\%
and 0.07\% respectively --- their fluxes are heavily suppressed by
cosmological dimming. These fractions were based on the assumption that
$L^{*}/L^{*}_{{\rm H}\alpha} = (0.2,0.3,0.5)$ for [SIII], [OIII] and [OII],
respectively.

We also considered an alternate approach to BOP using the ultraviolet rather
than the near-infrared and targeting Ly$\alpha$ emitters. We constructed an
empirical model of their line luminosity function which matches the observed
UV continuum luminosity function and Ly$\alpha$ equivalent width distribution
of $z \sim 3$ Lyman break galaxies \cite{Stei99,Shap03}. We found the UV
approach was not competitive --- the exposure times were an order of magnitude
larger. This is a combination of the much lower strength of the Ly$\alpha$
line and the fact that the background now {\em increases\/} towards higher
redshifts (0--3).

\section{Baryon Oscillations and Dark Energy constraints}

Of course there many possible science experiments BOP could address (see
Conclusions). We choose to focus on one for illustration of the redshift
survey power: detection of the baryon oscillations in the galaxy power
spectrum and resulting constraints on dark energy.

Based on the field of view and exposure time we calculate BOP can redshift
survey 2000 deg$^2$ and $10^7$ objects per year. We assume our two filter
optimized design and that each field is observed at two roll angles to improve
deconvolution of spectral overlap.  (This is also desirable to allow the zero
point of the wavelength scale of each object to be determined, alternatively
zeroth order could be used.)

The spectral resolution requirement remains to be determined precisely.  Blake
\& Glazebrook \cite{BG03} determined for full $P(k)$ resolution: $\sigma_z /
(1+z) \simeq 0.001$, which translates to $R=125$ if the line can be centroided
to a quarter of a resolution element at $S/N=10$. However to resolve the
nearby \Ha\ and [NII] lines requires $R>400$ so this might be preferable.

\begin{figure}[htbp]
\begin{center}
  \includegraphics[width=13cm]{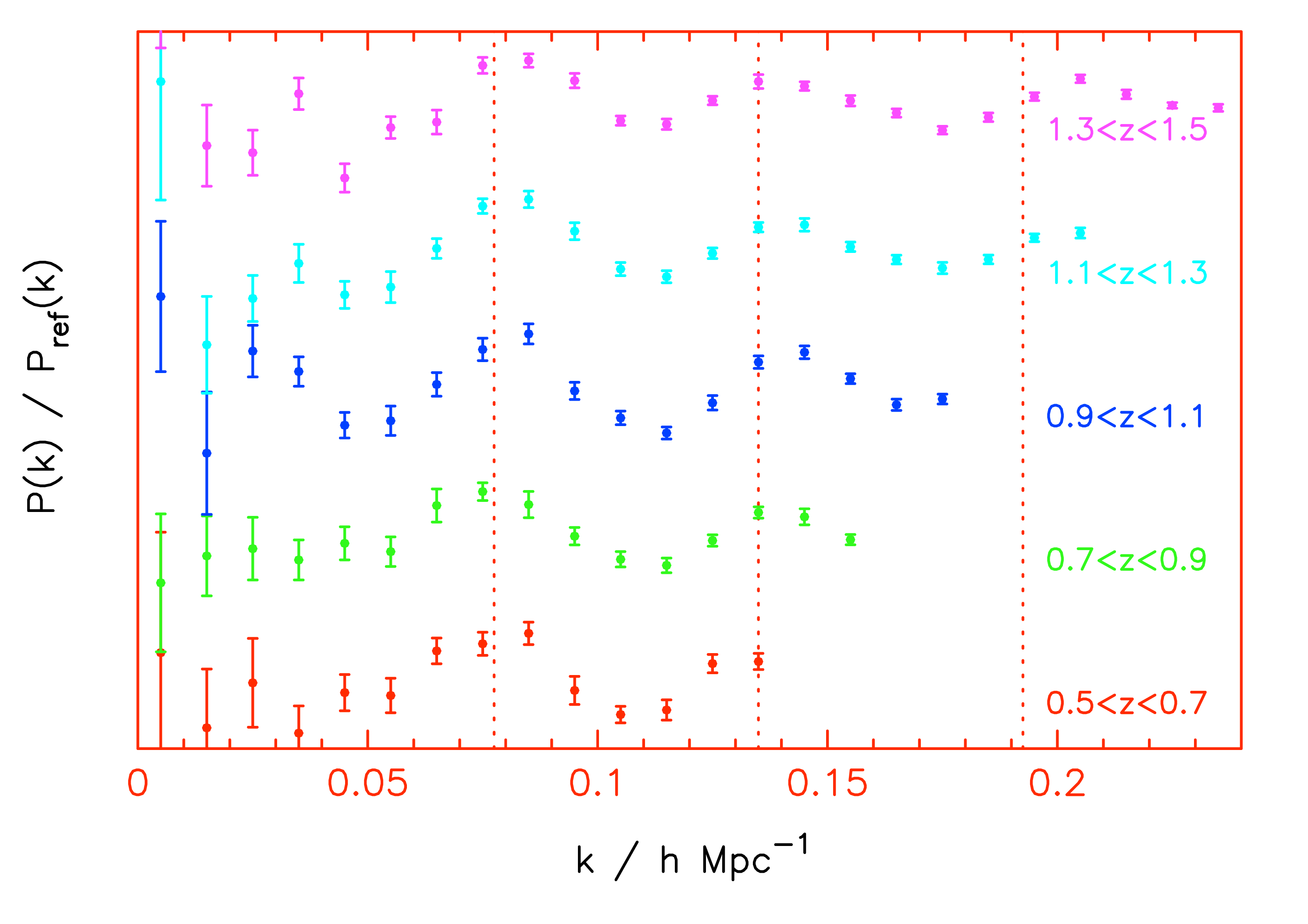}
  \caption{Predicted  galaxy power spectrum for a 10,000 deg$^2$ survey. 
    The non-oscillatory component has been divided out and the power spectrum (linear regime only)
    is shown in redshift bins. The small error bars show the precise
    measurement possible on the baryonic oscillations (the `standard ruler'
    for the dark energy test) possible with a survey of this size. In this
    case the model data is for a $w=-0.9$ simulated cosmology, the peaks
    visibly shift from their location in a $w=-1$ cosmology (vertical dotted
    lines).}
\label{fig:pk}
\end{center}
\end{figure}

\begin{figure}[htbp]
\begin{center}
  \includegraphics[width=13cm]{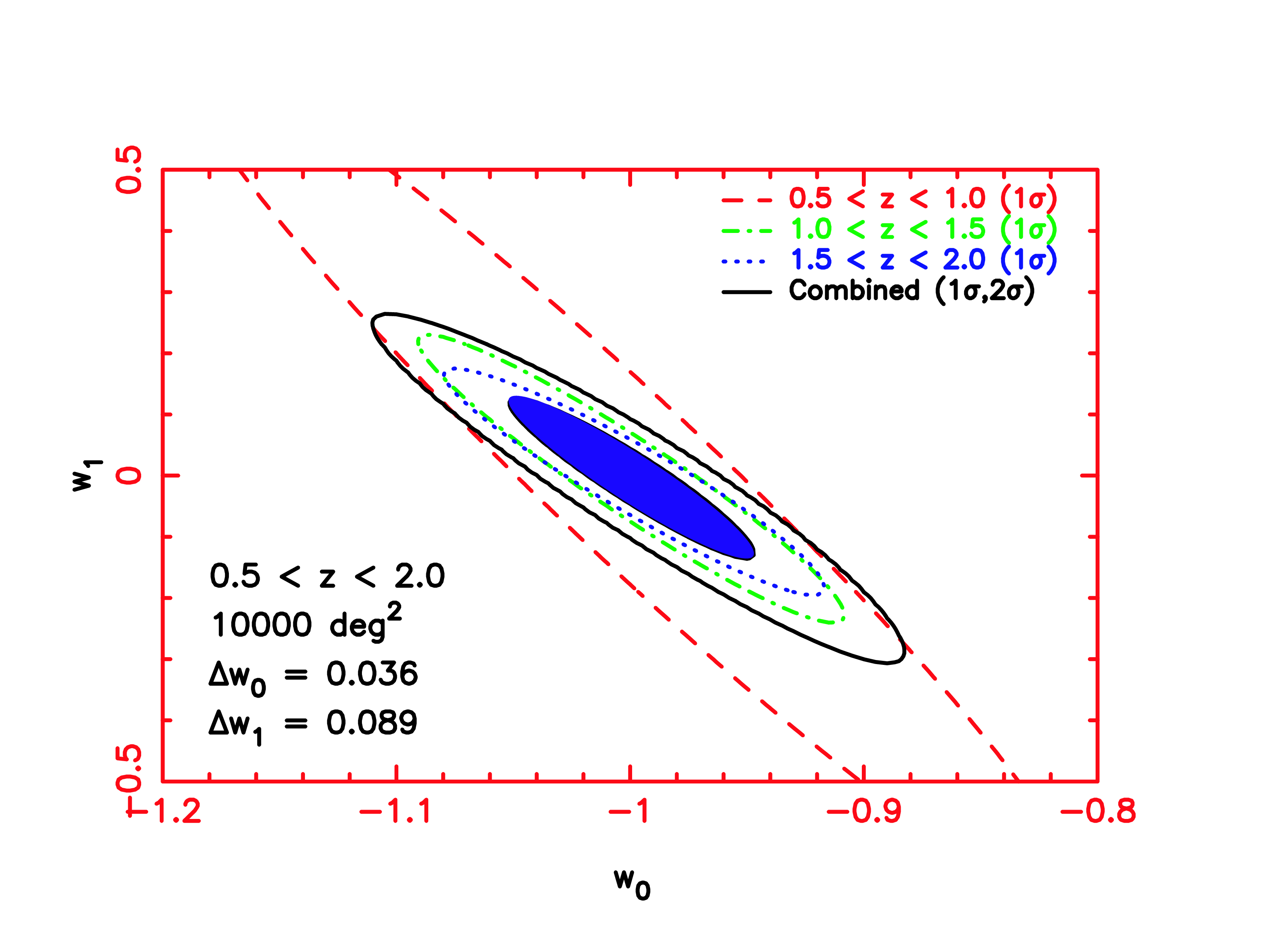}
  \caption{Recovered constraints on a dark energy equation of state
    $w(z) = w_0 + w_1 z$ from a 10,000 deg$^2$ BOP survey. It can be seen that
    each redshift range is slightly non-degenerate with the others, the
    combined constraints are very tight and are comparable to the precision
    delivered using supernovae in the SNAP mission.}
\label{fig:wz}
\end{center}
\end{figure}

We assume a 5 year mission resulting in a survey of 10,000 deg$^2$ and apply
the methodology of Glazebrook \& Blake (2004; in prep.) to simulate the
recovered power spectrum (Figure~\ref{fig:pk}) and resulting dark energy
constraints (Figure~\ref{fig:wz}). The constraints are comparable to those
delivered using the supernova technique by the Supernova Acceleration Probe
(SNAP \cite{SNAP}). As with the supernovae this requires independent accurate
constraints on other cosmological parameters (e.g. $\Omega_m$, $H_0$) from
local surveys and the Cosmic Microwave Background measurements --- we refer
the reader to Glazebrook \& Blake for a detailed discussion.

We note this proposed redshift survey improves by a factor of ten on the most
ambitious proposed future ground-based redshift survey. This is the KAOS
concept \cite{KAOS} for the Gemini Telescope which could survey 1000 deg$^2$.

It is also interesting to consider that there is very little conceptual
difference between BOP and {\em any} wide field imaging space mission. Both
require large detector arrays and wide field optics, the only difference is
provision of a grism. For example one could merge BOP with SNAP --- this would
add a completely independent method to the dark energy determination {\em and}
with similar precision. It would also mean that the many millions of
high-redshift galaxies imaged with SNAP could get spectroscopic follow-up in
the same mission. Alternately BOP is very similar to the DESTINY concept
\cite{DESTINY} which uses slitless spectroscopy to locate high-redshift
supernovae (albeit at somewhat lower resolution).

\section{Conclusions}

We have outlined a concept (BOP) for a near-infrared space mission using
slitless spectroscopy to do wide area redshift surveys an order of magnitude
larger than existing ground based surveys such as SDSS and 2dFGRS in both
number and volume and probing high-redshift ($0.5<z<2$). It represents a factor of
ten improvement on what could be done on the ground. Instrumentally it
represents a very simple hardware system which by taking advantage of the low
background in space has no requirement for complex fiber or slit configuring
machinery.   Although lower backgrounds could be achieved with wide 
field spectroscopic fiber or slit designs that have recently been 
proposed\cite{Mcc04} these technologies are 
immature by spaceflight standards and as such must be considered risky at the 
present time.

BOP would serve diverse science goals: some other examples include studying the
evolution of galaxy clustering, measuring the evolving luminosity function of
star-forming galaxies as a function of metallicity and environment, providing
redshift depth information for deep wide imaging surveys and searching for
Ly$\alpha$ emission from young galaxies at $z>7$. We have presented in
detail BOP's ability to constrain dark energy using baryon oscillation
measurements. It would deliver precision on the equation of state comparable
to the SNAP mission but via a completely independent technique.

The BOP concept is generalizable to any wide-field space mission as long as a
grism could be inserted in to the imaging system.  We argue that in any such
space mission the science-to-cost ratio of adding such a grism is enormous,
and should be considered.

\begin{ack}
  We would like to thank Chris Blake for assistance in
  producing Figures~\ref{fig:pk} and \ref{fig:wz} and Bob Woodruff from Lockheed Martin for useful design
  discussions. KG acknowledges generous funding from the David and Lucille
  Packard foundation.
\end{ack}

\end{document}